\documentclass[prl,twocolumn,showpacs,preprintnumbers,amsmath,amssymb]{revtex4}

\usepackage{amsmath} 
\usepackage{amssymb}
\usepackage{graphicx} 
\usepackage{array}
\usepackage{epstopdf}
\usepackage{xspace}

\newcommand{\BiSe}{Bi$_2$Se$_3$\xspace}

\newcommand{\ie}{{\it i.e.}}
\newcommand{\eg}{{\it e.g.}}

\begin{document}

\title{Ambipolar surface state transport in non-metallic stoichiometric Bi$_2$Se$_3$ crystals}

\author{Paul Syers}
 \affiliation{Center for Nanophysics and Advanced Materials, Department of Physics, University of Maryland, College Park, MD 20742}
\author{Johnpierre Paglione}
 \email{paglione@umd.edu}
 \affiliation{Center for Nanophysics and Advanced Materials, Department of Physics, University of Maryland, College Park, MD 20742}

\date{\today}


\begin{abstract}

Achieving true bulk insulating behavior in Bi$_2$Se$_3$, the archetypal topological insulator with a simplistic one-band electronic structure and sizable band gap, has been prohibited by a well-known self-doping effect caused by selenium vacancies, whose extra electrons shift the chemical potential into the bulk conduction band.
We report a new synthesis method for achieving stoichiometric Bi$_2$Se$_3$ crystals that exhibit nonmetallic behavior in electrical transport down to low temperatures.
Hall effect measurements indicate the presence of both electron- and hole-like carriers, with the latter identified with surface state conduction and the achievement of ambipolar transport in bulk Bi$_2$Se$_3$ crystals without gating techniques. 
With carrier mobilities surpassing the highest values yet reported for topological surface states in this material, the achievement of ambipolar transport via upward band bending is found to provide a key method to advancing the potential of this material for future study and applications.

\end{abstract}
\maketitle


The development of topological insulator (TI) materials has found rapid progress in the past few years \cite{Hasan82}. 
Distinguished from ordinary insulators by the so-called Z$_2$ topological invariants associated with the bulk electronic band structure \cite{Kane95,Fu74}, this class of materials is characterized by nonlocal topology of the electronic structure that gives rise to new electronic states with promise for realizing new technologies such as fault-tolerant quantum computation \cite{Nayak80}.  
By far the most widely studied system within the field of TI research is Bi$_2$T$_3$ (T=Se,Te) \cite{Xia5,Zhang5,Chen325,Chen329,Liu8,Deorani90}. 
To date, the major experimental efforts on these non-interacting bismuth-based TI materials have focused on refining measurement techniques in order to detect signatures of surface states. 
However, a continuing problem with the stoichiometric materials lies in the fact that they are not bulk insulators as predicted, but rather doped semiconductors \cite{Kohler71}. 
Both bulk and surface quality of TI materials are known to dramatically affect their properties, with the effects of site exchange (\eg, in Bi$_2$Te$_3$) or Se vacancy doping (\eg, in \BiSe) serving to introduce excess charge carriers in the bulk ($n$-type with \BiSe and $p$-type with Bi$_2$Te$_3$), reduce surface carrier mobilities and mix bulk and surface state conduction contributions. 

The common method of crystal growth using excess selenium falls short of reaching even a non-metallic temperature dependence of resistivity \cite{Butch81}, which has to date produced samples with some of the lowest bulk carrier concentrations ever reported.
Extensive work has been carried out to suppress bulk conductivity contributions by compensation doping \cite{Chen325,Chen329,Hor72,Hor79,Hsieh460} but this has only been achieved in the binary materials by introducing excess impurity scattering via chemical substitution methods \cite{Checkelsky103,Analytis6}, such as was accomplished using Se-Te site substitution in the case of the ternary compound Bi$_2$Te$_2$Se \cite{Ren82,Barreto14}.
Synthesis of defect-free epitaxial thin films \cite{Li22,Brahlek99} has also succeeded in reducing conduction through the bulk, and electrostatic gating techniques have been used to lower the chemical potential ($E_F$) into the bulk band gap regime \cite{Cho11,Kim8,Steinberg84}. But sensitivity to environmental conditions and crystalline quality \cite{Butch81} continue to pose problems for \BiSe.
This ultimately requires complicated and nuanced analysis of experimental data to identify and study the intrinsic nature of the topologically protected surface states, no matter the sample size; in the absence of further progress, increased attention is being devoted to other classes of materials \cite{Yan82,Dzero104,Lin9,Chadov9}. 

Here we employ a new bulk crystal growth technique to demonstrate the lowest attained bulk carrier concentrations in stoichiometric \BiSe, achieving a new regime of non-metallic transport behavior. Observations of a nonlinear Hall coefficient clearly identify the presence of two carrier types that can only be identified with separate bulk and surface state contributions to conductivity, with the coexistence of positive and negative carriers providing unequivocal proof of TI surface states from transport data alone.

Single crystals of \BiSe were grown under high gas pressures from ultrapure ($\geq$99.999\%) elemental Bi and Se via a self-flux technique \cite{Butch81}, utilizing a unique high pressure containment vessel (see further details of the growth procedure in Supplemental Information). The average crystal size varied from 0.5~mm~$\times$~1~mm to 3~mm~$\times$~3~mm.  Longitudinal and Hall resistance measurements were performed simultaneously on all samples reported here, using a six-wire configuration with two voltage contacts in standard longitudinal configuration and two voltage contacts in a transverse (Hall) configuration, both sharing the same current contacts.  All samples were measured in a commercial cryostat as a function of temperatures between 2 and 300 K and magnetic fields up to $\pm$14 T.

\begin{figure}[!t]
  \centering
  \includegraphics[width=3.4in]{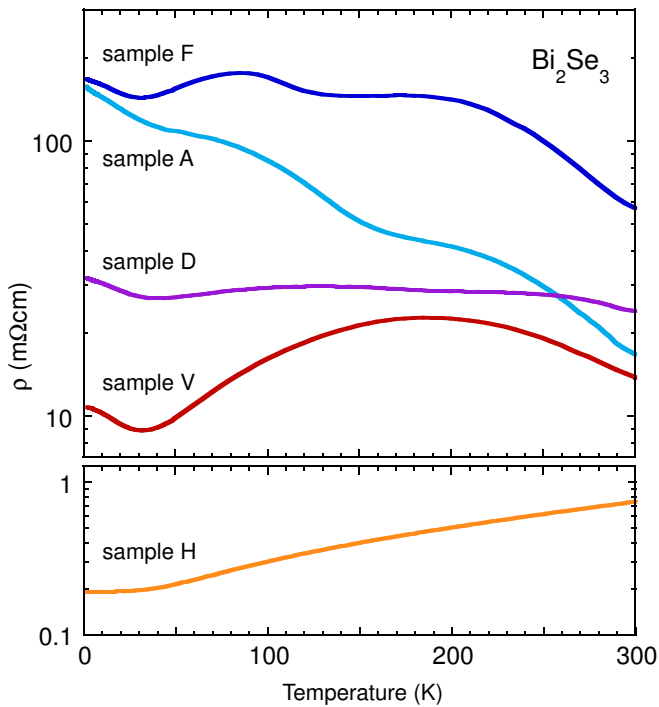}
  \caption{\label{RT} 
 Electrical resistivity of stoichiometric \BiSe crystals with varying carrier concentrations resulting from variations in sample growth conditions. Samples A, F, and D all show an overall increase in $\rho$ between 300 K and 2 K.  Samples V and H, shown for comparison, have semimetallic and metallic behavior.}
\end{figure}

The temperature dependence of the resistivity $\rho(T)$ shown in Fig.~\ref{RT} illustrates the range of non-metallic behavior of three samples of \BiSe crystals grown using the high-pressure technique. 
Two additional samples, V and H, are included for comparison. 
Sample H shows metallic behavior typical of most samples of \BiSe in the literature, while sample V shows semi-metallic behavior identical to other low carrier concentration pure samples reported to date \cite{Kohler71,Butch81}.  
Unlike samples V and H, the non-metallic crystals (A, D, F) exhibit an overall increase in resistivity with decreasing temperature. 
Furthermore, the most insulating-like samples exhibit a room temperature resistivity value far greater than the comparison samples or previous measurements of both pure and chemically substituted samples of Bi$_2$Se$_3$ \cite{Analytis6,Checkelsky103,Butch81}, indicating that the insulating behavior originates mainly from a clear decrease in overall carrier density (as opposed to a strong increase in scattering rate).
The presence of a distinct minimum in resistivity near 30~K in all samples follows the concentration-independent trend reported previously \cite{Kohler71,Butch81}, and is consistent with a phonon-dependent scattering feature \cite{Stordeur169} that only changes with lattice density such as induced by external pressure, which readily pushes the minimum up in temperature \cite{Hamlin}.
While Shubnikov-de Haas (SdH) oscillations can be discerned in low-temperature (2~K) magnetoresistance measurements of samples V and H, they are absent in the measurements of the non-metallic samples, indicating that the bulk carriers are at a low enough concentration to be in the quantum limit at moderate fields.

\begin{figure}[!t]
  \includegraphics[width=3.0in]{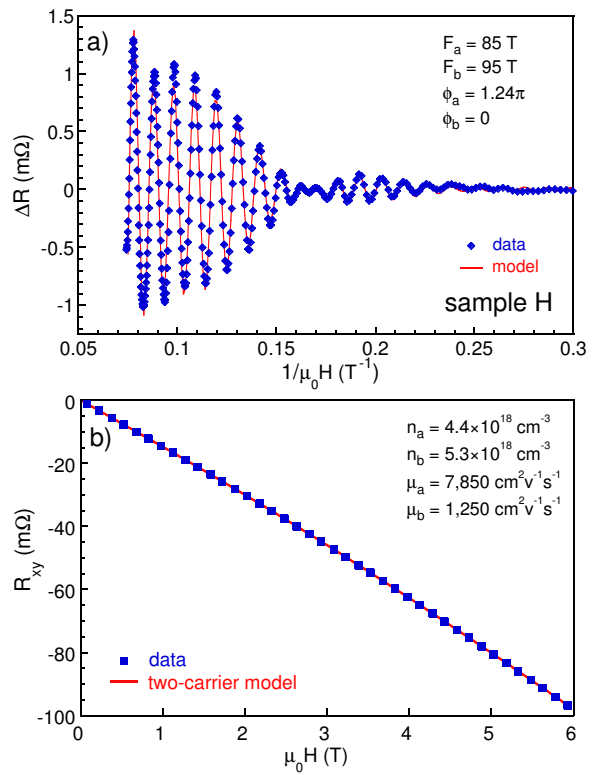}
  \caption{\label{Comparison} 
Analysis of quantum oscillations observed in \BiSe sample H, a high carrier density crystal. Panel a) presents the background-subtracted longitudinal resistance as a function of inverse magnetic field, with a fitted model (solid line) that incorporates the presence of two oscillation frequencies (85~T and 95~T) with a $1.24\pi$ phase shift between them. Panel b) presents a plot of the Hall resistivity for the same sample measured in-situ, along with a two-carrier model fit shown as a solid line. The resultant carrier densities and mobilities are shown in the inset, with carrier densities that match the values expected for the two oscillation frequencies noted above (see text for details).}
  \end{figure}

\begin{figure*}[!t]
  \includegraphics[width=6.5in]{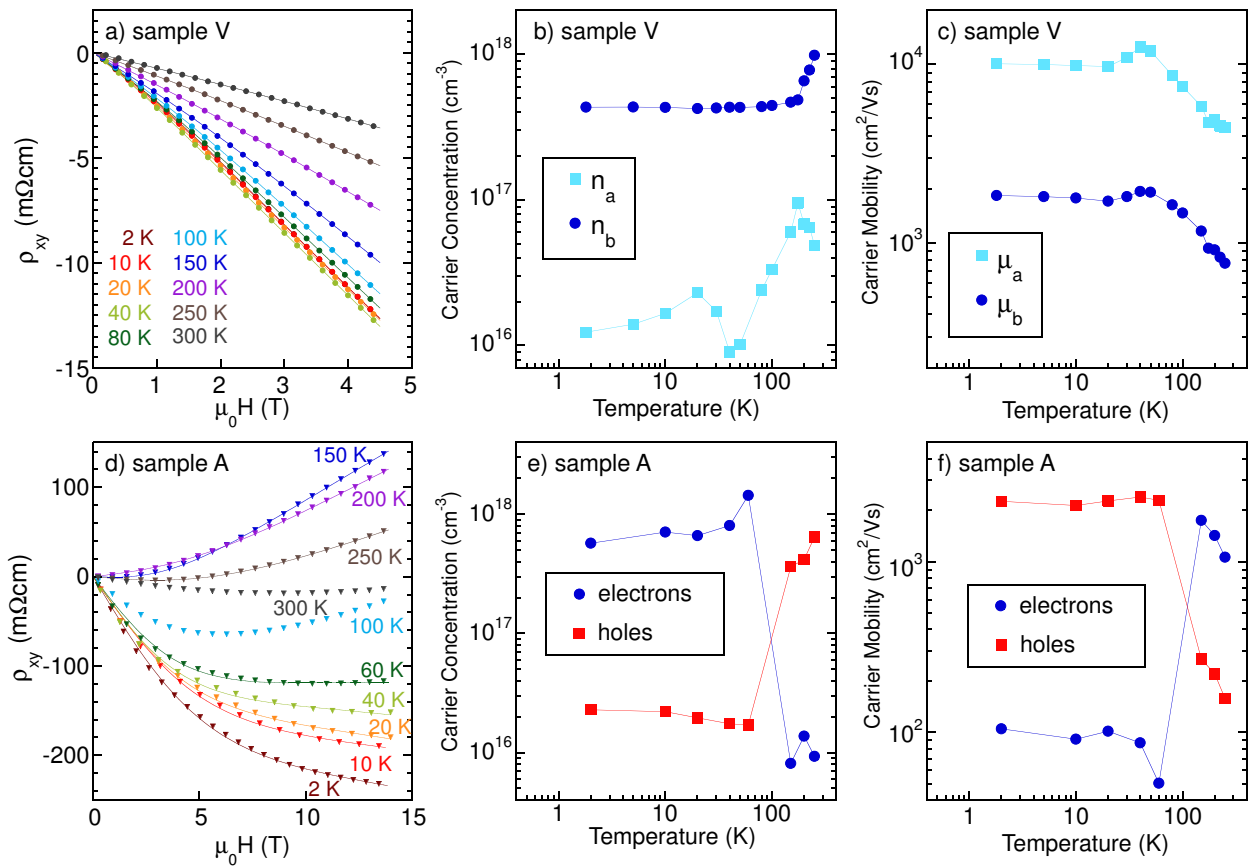}
  \caption{\label{Analysis} 
Hall effect data and analysis of single-crystal \BiSe obtained for two characteristic samples V and A, shown in panels a)-c) and d)-f), respectively. Transverse Hall resistance is presented in panels a) and d), with Drude model fits (see text) shown as solid lines. Panels b) and e) present the carrier densities extracted from the two-carrier analyis, and panels c) and f) present the resultant mobilities for each carrier type. The presence of two carrier contributions is easily discerned by the nonlinear behavior of $\rho_{xy}(H)$, in particular for sample A, which presents a crossover from electron- to hole-dominated conduction as a function of temperature. As described in the text, the two carrier types are ascribed to bulk and surface state carriers present in each sample, with hole-like conduction necessarily originating from surface states in sample A..}
\end{figure*}


A two-carrier Drude model was used to fit the Hall effect data for all samples, assuming two carriers of the same sign (electron-like) for samples D,V, and H and two with different sign carriers (one electron- and one hole-like) for samples A and F, respectively. 
Together with self-consistent fits to low-field ($\leq 1$ T) longitudinal magnetoresistance (see SI), a best match for the four physical parameters (density and mobility for each carrier) was reached for each of the non-metallic samples. 
To independently verify the fitting procedure, we first compare such results to analysis of SdH oscillations observable in higher carrier density samples. Samples V and H both exhibit SdH oscillations as usual for moderately doped samples \cite{Butch81}, but the latter sample exhibits a rare case of two oscillatory components, as clearly observable in the beating modulation presented in Fig.~\ref{Comparison}a). Fourier transform analysis confirms two oscillation frequencies of 85 T and 95 T, corresponding to 3D carrier densities of 4.4$\times10^{18}$~cm$^{-3}$ and 5.3$\times10^{18}$~cm$^{-3}$, respectively. This compares perfectly with the concentrations extracted from Hall data shown in Fig.~\ref{Comparison}b). 

Applying the standard Lifschitz-Kosevich formalism with an assumed typical effective mass of $0.1m_e$ \cite{Butch81}, we model the oscillations of samples V and H and extract Dingle temperatures and Berry's phase ($\phi$) information for the oscillations in both samples. 
From the analysis, carrier mobilities also compare favorably between SdH and Drude transport fit results (see SI for details). 
More surprising, for both samples the extracted $\phi$ values for the two oscillatory components are considerably offset (by nearly $\pi$) from one another, suggesting that the component with a non-zero Berry's phase is a TI surface state. 
This is extraordinary, considering that quantum oscillations of surface states in \BiSe have so far only been observed in very high (pulsed) magnetic fields in samples with enhanced bulk scattering \cite{Analytis6}. 
In the case of sample H, the corresponding mobility is much enhanced for the component with non-zero Berry's phase, approaching a value of 7,850~cm$^2$V$^{-1}$s$^{-1}$ as obtained by Drude analysis.

Using the same Drude analysis, we now compare Hall effect measurements and extracted mobility and carrier densities of semimetallic sample V (Figs.~\ref{Analysis}a-c) to those of the much more non-metallic sample A (Figs.~\ref{Analysis}d-f), which does not exhibit any observable trace of quantum oscillations. 
As shown, the Hall resistivities ($\rho_{xy}$) exhibit very unusual behavior with respect to magnetic field and temperature, especially in light of the well-characterized, single-band electronic structure of \BiSe \cite{Xia5}. 
Hall data for sample V is nearly linear, but exhibits a small but pronounced curvature in $\rho_{xy}(H)$ indicative of the presence of more than one type of charge carrier. 
Sample A, with more pronounced non-metallic behavior in $\rho(T)$, also exhibits much more pronounced nonlinearity in $\rho_{xy}(H)$, even crossing over to a hole-like response as temperature is raised.

The extracted mobility $\mu$ and carrier concentration $n$ values for each carrier type are shown in Figs.~\ref{Analysis}c)-f). Surprisingly, the electron carrier density 
$n_e = 4.3\times 10^{17}$~cm$^{-3}$  of sample A at 2~K is found to be higher than both carrier concentrations extracted from sample V data. However, the corresponding electron mobility is found to be extremely low (Fig.~\ref{Analysis}f), accounting for the lack of observable quantum oscillations in this sample.
The high mobilities of the minor bands in both samples are what account for their signatures in the Hall curves. 
Interestingly, the comparable values of mobilities of the lower carrier bands (approaching 8,000~cm$^2$V$^{-1}$s$^{-1}$) with that found in sample H is intriguing and suggestive of similar scattering processes. Most important, such large values are notably higher than any reported in the literature thus far, including those of MBE-grown thin-films with atomically sharp epitaxial interfaces \cite{Li12,Richardella97,Edmonds8}. 

Given the extensive efforts to increase mobilities of surface carriers through such efforts, it is important to investigate the manner by which this is achieved and to understand why perfecting crystal quality and/or suppressing the bulk carrier concentration is not sufficient.
The most striking result of the Hall analysis is the clear evidence of two carrier types, and moreover, evidence for two carriers with opposite signs in select samples.  
P-type samples of \BiSe have been previously reported \cite{Kohler71}, but the recent extensive set of measurements \cite{Xia5,Jozwiak84,Analytis81,Kong5,Zhu110,Wang109,Wray6,Zhu107,King107} studying the electronic structure of \BiSe have verified that its band structure is simplistic and includes only one bulk conduction and valence band together with Dirac surface states that cross the insulating gap. 
Therefore the most likely origin of the two carrier types is from bulk- and surface-derived bands. For two electron-like carriers (as for sample V), contributions from bulk and surface bands are understandable but one must also consider other causes, such as spin-split bulk bands \cite{Zhu107,King107} and trapped quantum well states due to downward band bending at the surface of the crystal \cite{Bianchi1,Bahramy3,Duan4}. 
However, the observation of hole-like carriers uniquely rules out such situations and allows for only one explanation: upward band bending. {\it This is the key aspect of achieving high surface state mobility.}

\begin{figure}[!t]
  \centering
  \includegraphics[width=3.25in]{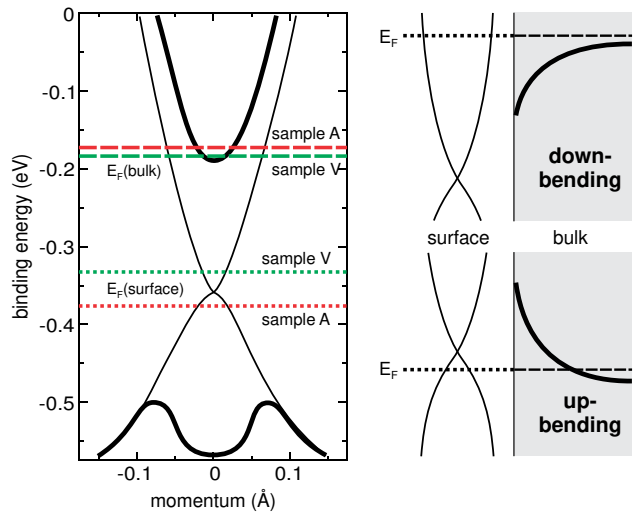}
  \caption{\label{Band} 
  (a) Electronic band structure of \BiSe\ obtained from photoemission experiments \cite{Bianchi1}, showing the positioning of bulk (dashed lines) and surface (dotted lines) chemical potential values $E_F$ extracted from two-carrier analysis of data for samples A and V (see text). Panel b) presents a schematic of the difference between downward and upward band bending near the surface, providing an explanation for the positioning of surface chemical potentials of samples V and A near the Dirac point, consistent with an upward band bending picture.}
\end{figure}

Using carrier densities estimated from Hall data analysis to calculate the corresponding two- and three-dimensional Fermi momenta $k_F$ for each sample, we map the positions of the surface and bulk chemical potentials onto the band structure measured by photoemission\cite{Bianchi27} as shown in Fig.~\ref{Band}. 
In line with previous studies \cite{Butch81,Kohler71}, the bulk band $E_F$ always appears to remain pinned to the edge of the conduction band and cannot be pushed into the gap for bulk samples by growth techniques alone.  
However, as shown the surface $E_F$ energies are distributed over a wider range, and even traverse the Dirac point. In our study, they do tend to remain close to the Dirac point, which may result from charge puddling \cite{Kim8} that acts to pin the surface $E_F$ there. The energy spacing between surface and bulk $E_F$ values indicate band bending of up to 190~meV, with the stronger band bending occuring in the more insulating samples and, most important, resulting in a hole-type carrier contribution to total conduction.
While the direction of the band bending at the surface of \BiSe is almost universally reported to be downward, most studies have been performed either on thin film samples or high carrier density samples.  One previous study performed on lower carrier density samples (\ie, comparable to sample V) has indeed reported upward band bending \cite{Analytis81}, possibly arising from surface-based interactions with elemental selenium. The suppression of selenium vapor pressure by the pressurized inert gas environment used in our crystal growth method may invoke a similar mechanism.

In any case, an upward band bending that places the surface chemical potential near the Dirac point yields the only plausible scenario that provides hole-type carriers in this band structure. Controlling the level of bending by growth tuning provides the ability to tune the position of the surface potential to be either above or below the Dirac point, demonstrating ambipolar transport of the Dirac surface states in large single crystals; an effect previously only achieved via gating techniques applied to ultra-thin films or samples \cite{Kim8}.
The significant enhancement in the measured TI surface state mobilities in stoichiometric \BiSe is surprising in comparison to prior extensive work on this material, and points to the importance of the unique material preparation technique that yields the uncommon band bending effect.
This is confirmed by our observations of changes in the transport data as a function of time (see SI). The suppression of resistivity values with air exposure time, in particular in the most insulating-like samples that exhibit hole-type behavior, is consistent with a significant downward shift in the energy bands at the surface, in agreement with previous studies of the electronic structure evolution at the surface of Bi$_2$Se$_3$ \cite{Mann4,Kong5}. 
Furthermore, the model of Se buildup at the surface of samples causing upward band bending is supported by findings that the carrier concentrations in samples increase with mechanical exfoliation \cite{Aguilar113,Kim8}.

Overall, while complete bulk insulating behavior in stoichiometric \BiSe remains difficult to achieve, our observations of greatly enhanced mobilities and ambipolar transport without atomically perfect thin films or fabricated gate structures suggests that engineering of electronic band bending near the surface of crystals via new routes of materials synthesis and preparation promises a new route to optimizing use of the simplest three-dimensional topological insulator.

\subsection{Acknowledgements}
We thank N.P. Butch, M. Fuhrer, and P.D.C. King for useful discussions, and J. Thornton for assistance pertaining to mathematical methods.
This work was supported by the Gordon and Betty Moore Foundation's EPiQS Initiative
through Grant GBMF4419.


\bibliography{Bi2Se3_Insulating_final}{}
\bibliographystyle{apsrev4-1}

\end{document}